\documentclass[12pt]{revtex4}
\usepackage{graphicx}
\usepackage{bm}
\usepackage{epsfig}

\begin{document}

\title{Bubble dynamics in double stranded DNA : A Rouse chain based approach}
\author{Rajarshi Chakrabarti }
\address{Department of Materials Science and Engineering, University of Illinois at Urbana-Champaign, USA}
\date{\today}

\begin{abstract}

We propose a model for the fluctuation dynamics of the local
denaturation zones  (bubbles)  in double-stranded DNA. In our
formulation, the DNA strand is model as a one dimensional Rouse
chain confined at both the ends. The bubble is formed when the transverse displacement of the chain attains a
critical value. This simple model effectively
reproduces the autocorrelation function for the tagged base pair in
the DNA strand as measured in the seminal single molecule experiment by
Altan-Bonnet et. al (Phys. Rev. Lett. 90, 138101 (2003)). Although our model is mathematically
similar to the one proposed by Chatterjee et al. (J. Chem. Phys. 127, 155104 (2007))  it goes beyond a single reaction coordinate description
by incorporating the chain dynamics through a confined Rouse chain and thus considers the collective nature of the dynamics. Our model also shows that the autocorrelation function is very sensitive to the relaxation times of the normal modes of the chain, which is obvious since the fluctuation dynamics of the bubble has the contribution from the different normal modes of the chain.

\end{abstract}

\maketitle
\date{\today}

\section{introduction}
In 1953 Watson and Crick \cite{Watson} proposed the structure of DNA
to be a stable double-stranded helix. The stability comes through
the staking interaction and the hydrogen bonding between the base
pairs in the opposite strands. But actually this picture represents
the equilibrium structure of DNA under physiological conditions. As
the interaction energy between these base pairs is only few $k_B T$,
even at room temperature due to thermal fluctuations locally DNA
strand opens up creating what is called ``bubbles". These bubbles
have different sizes and lifetimes. The creation and annihilation
kinetics of these bubbles is termed as the breathing dynamics. On
increasing the temperature or changing the pH these bubbles add up
to form larger bubbles and eventually the double stranded DNA
denatures.

In recent past single molecule fluorescence correlation spectroscopy
(FCS) experiment by Altan-Bonnet et. al \cite{Libchaber} has
provided the first quantitative insight to the relaxation kinetics
of the breathing mode of the double-stranded DNA. This breathing mode refers
to local denaturation and reclosing of the double-stranded structure. In their
experiment, two bases of the double-stranded DNA, corresponding to
opposite strands are tagged with a fluorophore and a quencher
respectively. So when the DNA structure is closed, fluorophore and
the quencher are in close proximity and the fluorescence is
quenched. But due to thermal fluctuation when the DNA structure
opens up creating a bubble, the fluorescence is restored. Hence the
base pair fluctuation leads to a fluctuation in fluorescence
intensity. This fluctuation in fluorescence intensity is monitored
by FCS which determines the characteristic dynamics of the
relaxation of dynamic correlations in the fluctuation of base pairs.
They introduced a correlation function

\begin{equation}
\label{gt}G(t)=\frac{\left< I(t)I(0)\right> -\left< I(t)\right>
\left< I(0)\right> }{\left< I(0)^2\right> -\left< I(0)\right> ^2}
\end{equation}

where $I(t)$ is the fluorescent intensity at time $t$. $G(t)$ was
found to be multiexponential. Interestingly the correlation function
for all the DNA constructs, at all temperatures follow the same
universal temporal behavior and when presented as a function of
rescaled time they all collapse into a single universal curve
$G(t/t_{1/2})$, where $t_{1/2}$ is such that $G(t/t_{1/2})=0.5$. To
explain the experimental data they proposed a simple kinetic model
in which $G(t/t_{1/2})$ becomes

\begin{equation}
\label{gexp}G_{ex}(z)=(1+\frac{m z}2)erfc(\sqrt{\frac{mz}2})-\sqrt{\frac{mz}%
\pi }e^{-\frac{mz}4}
\end{equation}

where, $z=\frac{t}{t_{1/2}}$ and $m$ is a parameter which is
adjusted to $0.328$ to ensure that $G_{ex}(1)=0.5$.

Since this experimental work by Altan-Bonnet et. al, the topic ``bubble dynamics"
as it is commonly refereed to has received a great deal of theoretical
attention. Just few years after this seminal experimental study of the transient time-dependent rupture
and re-healing of double stranded DNA, Bicout and Kats \cite{Bicout} proposed a kinetic scheme
based on two state model (closed or open) of the double stranded DNA.  Their formulation gave an analytical expression for
the survival probability, correlation function and life time for the bubble relaxation dynamics but formulation did not consider the
structure of the double stranded DNA at any level. Very recently Srivastava and Singh \cite{Srivastava} proposed a theoretical model where the interaction between the base pairs of the opposite DNA strands is described by Peyard-Bishop-Dauxois (PBD) \cite {Peyard} potential and the separation between the  base pairs (``y" in their notation)  follows a Fokker-Planck equation. Thus only making the interaction realistic but still not taking into account of the dynamics of the chain and restricting to a single relevant dynamical variable (separation ``y")  description. In a paper by Jeon, Sung and Ree \cite {Jeon} double stranded DNA was modeled as a duplex of
semiflexible chains mutually bonded by weak interactions in other words they used an extended worm-like chain model and a Langevin dynamics simulation was performed to examine the size distribution and dynamics of the bubble.
Shortly after this Fogedby and Metzler \cite{MetzlerPRL} came up with a theoretical model for the bubble dynamics
. They used the Poland-Scheraga free energy
\cite{Poland} where the free energy is a function of bubble size
$x$. The dynamics of $x$ follows a Langevin equation and the
corresponding Fokker-Planck equation is analogous to the imaginary
time Schrodinger equation for a particle moving in a Coulomb
potential subject to a centrifugal barrier. This mapping enabled
them to calculate the correlation function. But the best fit to the
long time behavior was obtained only when they considered the
dynamics of $x$ in a linear potential. Thus to reproduce the long
time data one can merely start with a linear potential but
unfortunately it may not fit to the experimental data in the
intermediate time range.   Interestingly the shortcomings
of the Fogedly-Metzler model was pointed out by Chatterjee et. al
\cite{Cherayil}. They assumed that the distance between fluorophore
and the quencher follows an overdamped Langevin equation in a
harmonic potential. Although their theory predicts the experimental
data reasonably well it does not take into account of the
fluctuation of different modes of the DNA strand contributing to the
bubble dynamics. A better theory should account for those and
actually our model does. In our formulation the DNA-strand is
described by a Rouse chain \cite{Doi1, Kawakatsu} confined at both
the ends. Transverse displacement of the string accounts for the
bubble formation. Naturally our model takes into account of the
contribution of different modes of the string to the breathing
dynamics.

\section{Our model}

We describe the bubble by a confined Rouse chain \cite{Doi1,
Kawakatsu} in a harmonic potential, $V(x)=\frac{1}{2}
\kappa R(n)^2$, where $R(n,t)$ denotes the position of the $n$th
segment of the confined Rouse chain in space and $t$ is the time. In
other words the chain is described by a field $R(n,t)$. Because of
the thermal fluctuations the bubble undergoes Brownian motion and
its time development is described by the equation

\begin{equation}
\label{rouseeq}\zeta \frac{\partial R(n,t)}{\partial
t}=k\frac{\partial ^2R(n,t)}{\partial n^2}-\kappa R(n,t)+f(n,t).
\end{equation}

In the above, $\zeta$ is the friction coefficient for the $n$th
segment and $\kappa$ is the force constant for the confining potential which
accounts for the staking interaction between the opposite strands of the DNA
molecule.

The fluorescent intensity at time $t$, denoted by $I(t)$ can be
written as

\begin{equation}
\label{it}I(t)=A\theta (R(a,t)-\alpha )
\end{equation}

where $R(a,t)$ denotes the position of the $a$th segment in space.
Later on we will choose $a$ such that it corresponds to the center
of the bubble. As mentioned earlier in the experiment one measures
the following correlation function. It is worth mentioning that this is the  the
relevant dynamical coordinate/variable in our model. Obviously it is not a one dimensional phenomenological reaction coordinate/dynamical variable like the separation between the donor and the acceptor as considered by Chatterjee et al. \cite{Cherayil} but actually a collective dynamical variable in the sense that it has the contribution from all the normal modes of the chain as is shown later in Eq.(6).

\begin{equation}
\label{gt}G(t)=\frac{\left< I(t)I(0)\right> -\left< I(t)\right>
\left< I(0)\right> }{\left< I(0)^2\right> -\left< I(0)\right> ^2}
\end{equation}

So in order to evaluate $G(t)$ we should first evaluate
$\left<I(t)I(0)\right>$ and $\left<I(t)\right>$. In our model the
correlation function, $\left<I(t)\right>$ would be

$$
\left<I(t)\right>=A\left<\theta (R(a,t)-\alpha )\right>
$$

Then we use the fact that the step function can be expressed as an
integral over delta function and in the next step we use the fourier
integral representation of the delta function.

$$
\left<I(t)\right>=\frac A{2\pi }\int\limits_\alpha ^\infty d\alpha
_1\int\limits_{-\infty }^\infty dk_1\left<e^{ik_1(R(a,t)-\alpha
_1)}\right>
$$

Now  $R(n,t)$ follows Rouse dynamics and it has a boundary
conditions $R(0,t)=0$ and $R(L,t)=0$, where $L$ is the chain length.
Keeping these boundary conditions in mind one can express $R(n,t)$
in terms of fourier modes

\begin{equation}
\label{rnt}R(n,t)=2\sum\limits_{p=1}^\infty X_p(t)\phi _p(n)
\end{equation}

$\phi(n)$ should satisfy the boundary conditions $\phi(0)=0$ and
$\phi(L)=0$. $X_p$ is the $p$th normal mode of the chain.

Substituting $R(n,t)$ from Eq. (\ref{rnt}) into the expression for
$\left<I(t)\right>$ to get

$$
\left<I(t)\right>=\frac A{2\pi }\int\limits_{-\alpha }^\infty
d\alpha _1\int\limits_{-\infty }^\infty dk_1e^{-ik_1\alpha
_1}\left<e^{2ik_1\sum\limits_{p=1}^\infty X_p(t)\phi _p(a)}\right>
$$

The above average is evaluated as follows

First we introduce $$b(s)=2k_1 \phi_p(a)\delta (t-s).$$ Then the
quantity in the angular bracket can be written as

$$
\left< e^{2ik_1\sum\limits_{p=1}^\infty X_p(t)\phi
_p(n)}\right\rangle =\left\langle e^{i\sum\limits_{p=1}^\infty
\int\limits_{-\infty }^\infty dsX_p(s)b(s)}\right>
$$

Then we use the definition of the characteristic functional
\cite{KuboBook}  and write the above quantity as

$$
\left< e^{i\sum\limits_{p=1}^\infty \int\limits_{-\infty }^\infty
dsX_p(s)b(s)}\right> =e^{-\frac 12\int\limits_{-\infty }^\infty
dt_1\int\limits_{-\infty }^\infty dt_2\sum\limits_{p=1}^\infty
b(t_1)H^{-1}(t_1-t_2)b(t_2)}
$$

where

$$
H^{-1}(t_1-t_2)=\left< X_p(t_1)X_p(t_2)\right>
$$

With this the integral in the exponent becomes

$$
\int\limits_{-\infty }^\infty dt_1\int\limits_{-\infty }^\infty
dt_2\sum\limits_{p=1}^\infty b(t_1)H^{-1}(t_1-t_2)b(t_2)=4k_1^2C(t)
$$

where

$$
C(0)=\sum\limits_{p=1}^\infty \phi _p(a)^2\left<
X_p(t)^2\right>=\sum\limits_{p=1}^\infty \phi _p(a)^2\left<
X_p(0)^2\right>
$$

 where we have used the fact $\left\langle X_p(t)^2\right\rangle =\left\langle
X_p(0)^2\right\rangle.$ Then

$$
\left\langle e^{i\sum\limits_{p=1}^\infty \int\limits_{-\infty
}^\infty dsX_p(s)b(s)}\right\rangle =e^{-2k_1^2C(0)}
$$

and subsequently $\left<I(t)\right>$ can be evaluated by integrating
over $\alpha_1$ and $k_1$.

$$
\left< I(t)\right> =\frac A{2\pi }\int\limits_\alpha ^\infty d\alpha
_1\int_{-\infty }^\infty dk_1e^{-ik_1\alpha _1-2k_1^2C(0)}=(\frac
A2)erfc[\frac \alpha {2\sqrt{2C(0)}}]
$$

Now the above expression for $\left<I(t)\right>$ does not include
the contribution coming from the displacement of the bubble in the
opposite direction, i.e. for negative values of $\alpha$. To include
that one has to multiply the above expression by $2$ to get the
final correct expression for $\left<I(t)\right>$.

\begin{equation}
\label{itfinal}\left\langle I(t)\right\rangle =\left\langle I(0)\right\rangle =A erfc[\frac \alpha {2\sqrt{%
2C(0)}}].
\end{equation}

Next we evaluate the correlation function $\left<I(t)I(0)\right>$.
Here also we follow the same technique and write the correlation
function as

$$
\left< I(t)I(0)\right>=(\frac{A^2}{4\pi ^2})\int\limits_\alpha
^\infty d\alpha _1\int\limits_\alpha ^\infty d\alpha _2\int_{-\infty
}^\infty dk_1\int_{-\infty }^\infty dk_2\left<
e^{2ik_1\sum\limits_{p=1}^\infty X_p(t)\phi
_p(a)}e^{2ik_2\sum\limits_{p=1}^\infty X_p(0t)\phi _p(a)}\right>.
$$

Next introducing $h(s)=2 k_1 \phi_p(a)\delta(t-s)+ 2 k_2
\phi_p(a)\delta (s)$  one writes the average as

$$
\left< e^{2ik_1\sum\limits_{p=1}^\infty X_p(t)\phi
_p(a)}e^{2ik_2\sum\limits_{p=1}^\infty X_p(0t)\phi _p(a)}\right>
=\left< e^{i\sum\limits_{p=1}^\infty \int\limits_{-\infty }^\infty
dsX_p(s)h(s)}\right>
$$

Similarly the above average can be written as

$$
\left< e^{i\sum\limits_{p=1}^\infty \int\limits_{-\infty }^\infty
dsX_p(s)h(s)}\right> =e^{-\frac 12\int\limits_{-\infty }^\infty
dt_1\int\limits_{-\infty }^\infty dt_2\sum\limits_{p=1}^\infty
h(t_1)H^{-1}(t_1-t_2)h(t_2)}
$$

Then one gets

$$
\left< e^{i\sum\limits_{p=1}^\infty \int\limits_{-\infty }^\infty
dsX_p(s)h(s)}\right> =e^{-2(k_1^2+k_2^2)C(0)-2k_1k_2C(t)}
$$

where

\begin{equation}
\label{ctsum}C(t)=\sum\limits_{p=1}^\infty \phi _p(a)^2\left\langle
X_p(t)X_p(0)\right\rangle
\end{equation}

 To get the final result one has to perform integrations over
$k_1$, $k_2$, $\alpha_1$ and $\alpha_2$. Unfortunately one of the
integrals over $\alpha$ (say $\alpha_2$) can not be performed
analytically and has to be carried out numerically. Now as done
earlier in the calculation of $\left<I(t)\right>$  here also one
should consider the contribution due to the displacement of the
string in the opposite direction which corresponds to negative
$\alpha$. This means not only the above expression should be
multiplied by a factor of $2$ but also there would be two cross
terms $ \left\langle \theta (R(a,t)-\alpha )\theta (R(a,0)+\alpha
)\right\rangle $ and $ \left\langle \theta (R(a,t)+\alpha )\theta
(R(a,0)-\alpha )\right\rangle $. These cross terms would contribute
equally to the correlation function. Including all these
contributions, the final correct expression for the correlation
function becomes

\begin{equation}
\label{iti0final}\left\langle I(t)I(0)\right\rangle
=\frac{A^2}{2\sqrt{2\pi C(0)}}(a_1(t)+a_2(t))
\end{equation}

where

\begin{equation}
\label{a1}
a_1(t)=\int\limits_\alpha ^\infty d\alpha _2e^{-\frac{\alpha _2^2}{8C(0)}%
}erfc[\frac{2C(0)\alpha -C(t)\alpha
_2}{2\sqrt{2C(0)(4C(0)^2-C(t)^2)}}]
\end{equation}

and

\begin{equation}
\label{a2}
a_2(t)=\int\limits_{-\infty }^{-\alpha }d\alpha _2e^{-\frac{\alpha _2^2}{%
8C(0)}}erfc[\frac{2C(0)\alpha +C(t)\alpha _2}{2\sqrt{2C(0)(4C(0)^2-C(t)^2)}}%
]
\end{equation}

 Now in principle one can use Eq. (\ref{itfinal}) and Eq.
(\ref{iti0final}) to calculate $G(t)$ defined in Eq. (\ref{gt}). But
to do that one has to evaluate $C(t)$ first. Now as mentioned
earlier $\phi_p(n)$ introduced in Eq. (\ref{rnt}) has to vanish at
the boundaries which is satisfied by choosing
$\phi_p(n)=\sin(\frac{p\pi n}{L})$. One can show that the time
dependent coefficients $X_p(t)$ obeys the following equation.

\begin{equation}
\label{xpeq}\zeta _p\frac{\partial X_p(t)}{\partial
t}=-k_pX_p(t)+f_p(t).
\end{equation}

$$
\zeta _p=2L\zeta
$$
and
$$
k_p=(\frac{2k\pi ^2p^2}L+2\kappa L)
$$
and the $f_p(t)$'s are the random forces which satisfy

$$
\left\langle f_p(t)\right\rangle =0
$$

$$
\left\langle f_p(t)f_p(s)\right\rangle =2\zeta _pk_BT\delta (t-s)
$$

From Eq. (\ref{xpeq}) and using the statistical properties of the
random forces one can derive

\begin{equation}
\label{xpcorr}\left\langle X_p(t)X_p(0)\right\rangle =\frac{k_BT}{k_p}%
e^{-\frac t{\tau _p}}
\end{equation}
where

$$
\tau _p=\frac{\zeta _p}{k_p}=\frac{L^2\zeta }{k\pi ^2p^2+\kappa L^2}
$$

Next we will choose $a=\frac{L}{2}$ which corresponds to the mid
point of the bubble. This further reduces the expression for $C(t)$
to

\begin{equation}
\label{ctsum2}C(t)=k_BT\sum\limits_{p=odd}\frac{e^{-\frac{(k\pi
^2p^2+\kappa L^2)}{L^2\zeta }t}}{(\frac{2k\pi ^2p^2}L+2\kappa L)}
\end{equation}

Then using Eq. (\ref{xpcorr}) and replacing the sum in Eq.
(\ref{ctsum2}) by an integral one can write $C(t)$ as

\begin{equation}
\label{ctfinal1}C(t)=\frac{k_BT}2\int\limits_0^\infty dp\frac{e^{-\frac{%
(k\pi ^2p^2+\kappa L^2)}{L^2\zeta }t}}{(\frac{2k\pi ^2p^2}L+2\kappa
L)}
\end{equation}

The factor $\frac{1}{2}$ comes because only odd modes contribute to
the sum. Fortunately the above integral is analytical. After
carrying out the integration one gets

\begin{equation}
\label{ctfinal2}C(t)=\frac{k_BT}{8\sqrt{k\kappa }}erfc[\sqrt{\frac{\kappa t}%
\zeta }].
\end{equation}

Using this expression for $C(t)$ and Eq. (\ref{itfinal}) and Eq.
(\ref{iti0final}) one can evaluate $G(t)$ defined in Eq. (\ref{gt}).

\section{Comparison with Altan-Bonett Experiment}
Here we make a comparison of our model with the experimental data obtained by Altan-Bonett et al \cite{Libchaber}.The fitting function they used
has already been mentioned at the beginning ( Eq. (\ref{gexp})). Figure 1 is a comparison of our model with Altan-Bonett fitting function (Eq. (\ref{gexp})). The comparison are made for a fixed set of parameters ($\protect\alpha =1,L=10, k=100, \zeta=5, k_BT=1$) while changing the force constant of the confined harmonic well ($\kappa$) which is also a measure of the strength of H-bonding and staking interaction between the base pairs. Changing $\kappa$ by keeping other parameters unchanged physically means changing the relaxation time of the normal modes of the chain as $\tau _p=\frac{\zeta _p}{k_p}=\frac{L^2\zeta}{k\pi ^2p^2+\kappa L^2}$. Thus a smaller value of $\kappa$ results slow relaxation of the normal modes and a larger $\kappa$ results faster relaxation of normal modes. Another interesting observation is that the relaxation time for the higher normal modes (larger $p$) have very weak dependence on $\kappa$, while the relaxation times for the lower normal modes (smaller $p$) have stronger $\kappa$ dependence. One can see from Fig. 1 that there exist at least one value of $\kappa$ (when all the other parameters are kept fixed) for which a very good comparison with Altan-Bonett results can be made. Moreover a small change in the value of $\kappa$ results poor comparison with the experimental result as shown in Fig. 1 (dashed, blue). A smaller value of $\kappa$ makes the dynamics slower as expected and also a higher value results faster decay of the autocorrelation function (green, dashed-dot). This strong $\kappa$ dependence also suggests that the lower normal modes of the strand mostly contribute to bubble dynamics. Here we would also like to mention that the choice of a particular set of parameters is not unique as more than one set of values can also be used to make a reasonably well comparison as is also found with the model of Chatterjee et al \cite{Cherayil}.

\subsection{long and short time behavior of $G(t)$}

In this section we explore the short and long time behavior of
$G(t)$. Let us first consider the correlation function used to fit
the experimental data by Altan-Bonett Eq. (\ref{gexp}). At short
time one can approximate, $erfc[z]\simeq 1-\frac{2 z}{\sqrt{\pi}}$
as ($z\rightarrow 0$) and the correlation function becomes
$G_{ex}(z)=1-2\sqrt{\frac{1}{\pi z}}z^{1/2}$. Thus it behaves as a
power law. To explore the long time limit we use the following
approximation, $ erfc[z]\simeq \frac{e^{-z^2}}{z\sqrt{\pi }} $ as
($z\rightarrow \infty$), which gives $
G_{ex}(z\rightarrow \infty)=\frac{e^{-\frac{az}4}}{\sqrt{az}\sqrt{\pi }} $. Now it
would be interesting to see what happens to $G(t)$ (or $G(z)$,
$z=t/t_{1/2}$) in our formulation. All time dependence in $G(t)$
comes through $\left<I(t)I(0)\right>$ or in other words through
$a_1(t)$ and $a_2(t)$ defined earlier. To analyze the the
short time behavior of $G(t)$ we first rewrite short time $C(t)$ as

\begin{equation}
\label{ctfinal}C(t)=C(0)(1-2 c_1 \sqrt{t})
\end{equation}

 where, $c_1=\sqrt{\frac{\kappa}{\pi\zeta}} $  and  $C(0)=\frac{k_BT}{8\sqrt{k\kappa }}$.

 Using the above short time expression for $C(t)$ and keeping the leading order in $t$
 the short time expression of $a_1(t)$ simplifies to

\begin{equation}
\label{a1short} a_1(t)=\int\limits_\alpha ^\infty d\alpha _2e^{-\frac{\alpha _2^2}{8C(0)}}(A_1(\alpha_2)+B_1(\alpha_2)\sqrt{t})=K_1+L_1\sqrt{t}
\end{equation}

 with $A_1(\alpha_2)={\frac{1}{\sqrt{C(0)}}} (\sqrt{\frac{2}3}\alpha-\frac{\alpha_2}{\sqrt{6}})$ and $B_1(\alpha_2)=C(0)^{-\frac{3}2}({{\frac{2\sqrt{6}}9}}c_1\alpha_2-\frac{\sqrt{6}}9 c_1 \alpha).$

 Similarly

 \begin{equation}
\label{a2short} a_2(t)=\int\limits_{-\infty }^{-\alpha }d\alpha _2e^{-\frac{\alpha _2^2}{8C(0)}}(A_2(\alpha_2)+B_2(\alpha_2)\sqrt{t})=K_2+L_2\sqrt{t}
\end{equation}

where

 $A_2(\alpha_2)={\frac{1}{\sqrt{C(0)}}} (\sqrt{\frac{2}3}\alpha+\frac{\alpha_2}{\sqrt{6}})$, $B_2(\alpha_2)=C(0)^{-\frac{3}2}(-{{\frac{2\sqrt{6}}9}}c_1\alpha_2-\frac{\sqrt{6}}9 c_1 \alpha)$,
 $K_1=\int\limits_\alpha ^\infty d\alpha _2e^{-\frac{\alpha _2^2}{8C(0)}}A_1(\alpha_2)$, $L_1=\int\limits_\alpha ^\infty d\alpha _2e^{-\frac{\alpha _2^2}{8C(0)}}B_1(\alpha_2)$, $K_2=\int\limits_\alpha ^\infty d\alpha _2e^{-\frac{\alpha _2^2}{8C(0)}}A_2(\alpha_2)$, $L_2=\int\limits_\alpha ^\infty d\alpha _2e^{-\frac{\alpha _2^2}{8C(0)}}B_2(\alpha_2)$.

Then combining  Eq. (\ref{a1short}),  Eq. (\ref{a2short}) one finally gets the short time limit of $G(t)$.

\begin{equation}
\label{shortgt} G(t)=G(0)(1-\lambda\sqrt{t})
\end{equation}

where, $\lambda=\frac{d_1 (L_1+L_2)}{(d_1 K_1+d_1K_2-d_2)}$, $G(0)=1$ (as $G(t)$ is normalized), $d_1=(erfc[\frac \alpha {2\sqrt{%
2C(0)}}])^2$, $d_2=\frac{1}{2\sqrt{2C(0)}}$.

Hence at short time it has the similar time
dependence as the one used by Altan-Bonett to fit the experimental
data. Next we analyze the long time behavior of $G(t)$. As mentioned earlier
all the time dependence of $G(t)$ comes from $C(t)$ which is embedded in $a_1(t)$ and $a_2(t)$. As in the long
time, $C(t)<<C(0)$, one can further simplifies the expressions for $a_1(t)$ and $a_2(t)$. To do this consider the complimentary
error function sitting inside the integral in Eq. (\ref{a1}).

$$erfc[\frac{2C(0)\alpha -C(t)\alpha
_2}{2\sqrt{2C(0)(4C(0)^2-C(t)^2)}}]=erfc[\frac{2C(0)\alpha -C(t)\alpha
_2}{2\sqrt{2C(0)(4C(0)^2)}}].$$

Now the complimentary error function is in the form $erfc[P_1-Q_1 C(t)]$ with $P_1=\frac{\alpha}{2\sqrt{2 C(0)}}$, $Q_1=\frac{\alpha_2}{{4\sqrt{2}}{(C(0)^{-3/2})}}$.As $C(t)$ approaches zero in the long time we can make a series expansion of the above complimentary error function and keep only the first two terms to get $$erfc[P_1-Q_1 C(t)]=erfc[P_1]+\frac{2 e^{-P_1^2} Q_1 \sqrt{t}}{\sqrt{\pi}}.$$ Now one can analytically perform the integration over $\alpha_2$ to get the long time expressions for $a_1(t)$. Similarly we get the long time limit of $a_2(t)$ by following the above steps. It shows that the autocorrelation function $G(t)$ approaches zero the same way $C(t)$ approaches zero in the long time limit. Thus in the long time limit $G(t)$ approaches zero as $\frac{e^{-t}}{\sqrt{t}}$ as is also predicted by Altan-Bonett fitting model.

\begin{figure}[tbp]
\centering
\epsfig{file=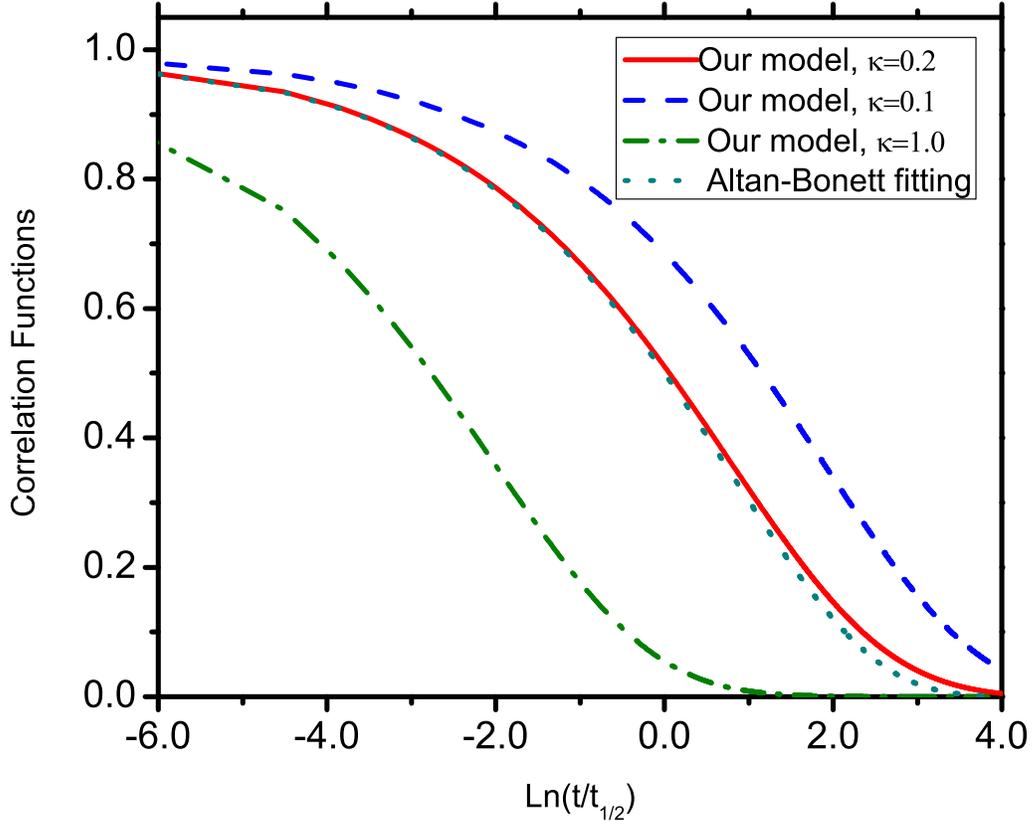,width=1\linewidth}\newline
\caption{The autocorrelation function against time ($t/t_{1/2}$).
 The values of other parameters used are: $\protect\alpha %
=1,L=10, k=100, \zeta=5, k_BT=1 $.} \label{correlation}
\end{figure}

\section{Conclusions}

The stochastic dynamics of DNA bubble formed due to rupture and reformation of hydrogen bonds is modeled based on a Rouse chain description
of the DNA strand. Although the model is very simple it produces the experimental results of Altan-Bonett reasonably well. Unlike other well known models \cite{MetzlerPRL, Cherayil} our model goes beyond a single reaction coordinate or order parameter description by taking into account of the collective nature of the dynamics through different modes of the chain as is done in the Rouse description in the simplest possible way. The dynamics seems to be very sensitive to the relaxation times of different normal modes of the chain which is physically understandable as the ``bubble dynamics" should have the contribution from all the possible normal mode of vibration of the chain. However the current model probably can not account for the bubble size distribution. We are in the process of developing a more realistic model which can account for the issue like bubble size distribution.

\section {acknowledgement}

The author thanks K. L. Sebastian for encouragements.

\bibliographystyle{apsrev}

\end{document}